\documentclass[conference]{IEEEtran}
\usepackage{array}
\usepackage{booktabs} 
\usepackage{multirow}
\usepackage{bm}

\usepackage{mathrsfs}
\usepackage{cite}
\usepackage{amsmath,amssymb,amsfonts}
\usepackage{graphicx}
\usepackage{textcomp}
\usepackage{xcolor}
\usepackage[T1]{fontenc}
\usepackage[utf8]{inputenc}
\usepackage[font=small,tableposition=top]{caption}
\usepackage{booktabs}
\usepackage{url}
\usepackage{algorithm}
\usepackage{algpseudocode}
\usepackage{amsmath}
\usepackage{verbatim}

\usepackage{caption}
\usepackage{tabularx}

\usepackage{nomencl}
\makenomenclature

\usepackage{graphicx}
\usepackage{subfig}

\usepackage{CJK}
\usepackage{type1cm}
\usepackage{times}
\usepackage[marginal]{footmisc}
\usepackage{stfloats}

\def\BibTeX{{\rm B\kern-.05em{\sc i\kern-.025em b}\kern-.08em
		T\kern-.1667em\lower.7ex\hbox{E}\kern-.125emX}}
\ifCLASSINFOpdf
\else
\fi
\begin{document}
%
\title{Fast Distributed Stochastic Scheduling for A Multi-Energy Industrial Park
\thanks{This work was supported by the National Key Research and Development Program of China
(2018YFB1702300) and the National Natural Science Foundation of China under Grant 61731012.}
}

\author{\IEEEauthorblockN{Dafeng Zhu}
\IEEEauthorblockA{Department of Automation \\
Shanghai Jiao Tong University\\
Shanghai, China \\
dafeng.zhu@sjtu.edu.cn}
\and
\IEEEauthorblockN{Bo Yang}
\IEEEauthorblockA{Department of Automation \\
Shanghai Jiao Tong University\\ 
Shanghai, China \\
bo.yang@sjtu.edu.cn}
\and
\IEEEauthorblockN{Zhaojian Wang}
\IEEEauthorblockA{Department of Automation \\
Shanghai Jiao Tong University\\
Shanghai, China \\
wangzhaojiantj@163.com}
\and
\IEEEauthorblockN{Chengbin Ma}
\IEEEauthorblockA{University of Michigan-\\Shanghai Jiao Tong University Joint Institute \\
Shanghai Jiao Tong University\\
Shanghai, China \\
chbma15@gmail.com}
\and
\IEEEauthorblockN{Kai Ma}
\IEEEauthorblockA{School of Electrical Engineering \\
Yanshan University\\
Qinhuangdao, China \\
kma@ysu.edu.cn}
\and
\IEEEauthorblockN{Shanying Zhu}
\IEEEauthorblockA{Department of Automation \\
Shanghai Jiao Tong University\\
Shanghai, China \\
shyzhu@sjtu.edu.cn}
}

\maketitle

\begin{abstract}
The multi-energy management framework of industrial parks advocates energy conversion and scheduling, which takes full advantage of the compensation and temporal availability of multiple energy. However, how to exploit elastic loads and compensate inelastic loads to match multiple generators and storage is still a key problem under the uncertainty of demand and supply. 
To solve the issue, the energy management problem is constructed as a stochastic optimization problem. The optimization aims are to minimize the time-averaged energy cost and improve the energy efficiency while respecting the energy constraints. To achieve the distributed implementation in real time without knowing any priori knowledge of underlying stochastic process, a distributed stochastic gradient algorithm based on dual decomposition and a fast scheme are proposed. 
 The numerical results based on real data show that the industrial park, by adopting the proposed algorithm, can achieve social welfare maximization asymptotically.

\end{abstract}

\begin{IEEEkeywords}
Multi-energy management framework, industrial park, distributed implementation, stochastic gradient algorithm
\end{IEEEkeywords}

%
\IEEEpeerreviewmaketitle

\footnotetext{This work was supported by the National Key Research and Development Program of China
(2018YFB1702300) and the National Natural Science Foundation of China under Grant 61731012.}

\section{Introduction}

With the increasing industrial production scale, energy consumption has grown rapidly, which is the main driving force of industrial parks to tackle the serious problems of low energy efficiency and increasing operating cost. To solve these issues, 
 multi-energy generation plants (MEGPs), including combined heat and power (CHP) units, photovoltaic panels, energy storages and boilers, are integrated into the industrial park. By shifting supply/demand across spatiotemporal scales among multi-energy networks, MEGPs can improve energy efficiency and income \cite{Gu2020Bi}. MEGPs can gain schedule complementarity and flexibility by control and optimization of multi-energy networks, which can achieve enhanced reliability, high energy utilization, and increased efficiency.

Many studies have been done on the multi-energy management of industrial parks. In \cite{Chen2020Improving}, a novel method of adopting the hydraulic inertia of steam heating network is proposed to improve the flexibility of multi-energy industrial parks.
 In \cite{Jiang2017Interaction}, a generalized multi-energy demand-response interaction optimization model is established to realize the interaction among users and the power grid in an industrial park. In \cite{Liu2020Heat}, a multi-energy industrial park's management framework is established, and the energy conversion and interaction between CHP unit owners and users are considered to realize the peak load shifting. 
 However, most of the existing studies mainly address the issues of interaction between demand response and supply side without considering the combination of multi-energy storage and stochastic nature of the scenario. 

Although renewable energy and energy storage is an effective way to release the unbalance of supply and demand, the time-varying and stochastic nature of renewable energy generation should be taken into account. 
 A large number of studies aim to solve the stochastic problem of energy management. In \cite{Deng2014Load}, dual decomposition and stochastic gradient are proposed to address the optimization problem through appropriate scheduling, that is, shifting the peak energy demand by pricing tariffs as incentives. 
 In \cite{Etesami2018Stochastic}, the interaction among prosumers is formulated as a stochastic game, and a novel distributed algorithm is proposed to achieve the optimal payoff. In \cite{Liu2018Energy}, a day-ahead scheduling model of an energy sharing provider is built to improve the power profile and increase the operating profit via stochastic programming. These works mainly concentrate on electricity scheduling and do not consider multi-energy complementary utilization. Moreover, the aforementioned works do not consider the convergence rate improvement of proposed algorithm, which is important for real-time implementation.


To solve the energy management problem,
 this paper presents a distributed stochastic gradient algorithm based on the dual decomposition and a fast scheme. 
The dual decomposition is applied to achieve distributed implementation with temporally-coupled constraints, and the fast scheme can ensure real-time implementation. 
The main contributions in this paper are listed as follows.
\begin{itemize}
\item
A multi-energy management framework is presented for an industrial park, where MEGPs supply energy to industrial users. Compared with other methods, this framework fully mobilizes the coordination mechanism of energy supply, elastic load (EL), inelastic load (IL) and storage with a more comprehensive model. 
\item 
The variation of multi-energy loads, renewables and energy prices is considered to construct a stochastic optimization problem. Instead of paying barely attention to the electricity storage, the paper also takes into account the heat storage, time-varying price and stochastic multi-energy demand and supply.
\item To reduce the influence of multi-energy coupling, stochastic demand and renewable energy generation, a distributed algorithm and a fast scheme are proposed. The fast scheme ensures real-time coordination of instantaneous scheduling.

\end{itemize}

The remainder of the work is listed as follows. The system model in the industrial park is introduced in Section \uppercase\expandafter{\romannumeral2}. Section \uppercase\expandafter{\romannumeral3} proposes the stochastic energy optimization and distributed realization methods. 
The simulation results based on real data are shown in Section \uppercase\expandafter{\romannumeral4}. Finally, the conclusion is given in Section \uppercase\expandafter{\romannumeral5}.

\section{System Model}
\begin{figure}
  \centering
  \includegraphics[width=.8\hsize]{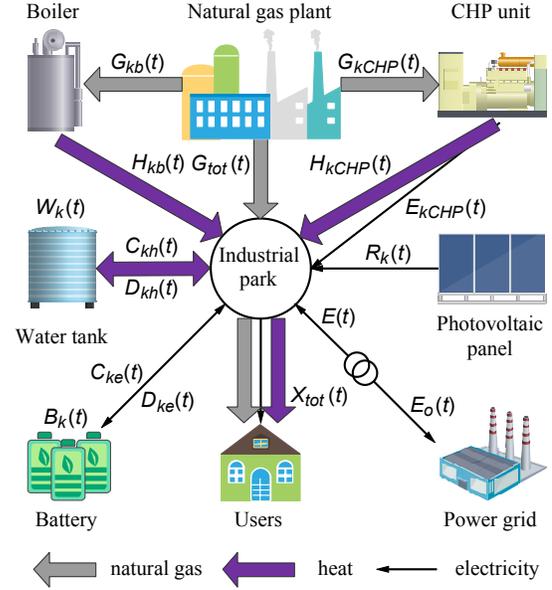}
  \caption{Energy flows of the industrial park}
  \label{fig1}
\end{figure}

This paper considers a system consisting of an industrial park, an electricity utility company, and a gas utility company, where three types of energies are supplied, i.e., electricity, heat and natural gas, as shown in Fig.~\ref{fig1}. The industrial park is composed of users  and MEGPs \cite{Yazdani2018Strategic}. The MEGPs include CHP units, photovoltaic panels, batteries, water tanks and boilers.
The park has $\boldsymbol{K}=\{1, 2, ..., K\}$ MEGPs. The energy devices in next subsection are modeled for MEGP $k$, $k\in \boldsymbol{K}$. 
Here, a time slot is one hour to coordinate with the simulation.

\subsection{Multi-energy Generation Plant}
In this subsection, the models of energy storage systems $k$, CHP unit $k$ and boiler $k$ in MEGP $k$ are introduced. 

\subsubsection{Energy Storage Systems}
 The electricity and heat storage models are denoted as 
\begin{equation}
B_{k}(t+1)=B_{k}(t)+\eta_{cke}C_{ke}(t)-\frac{1}{\eta_{dke}}D_{ke}(t)
\label{A1}
\end{equation}
\begin{equation}
W_{k}(t+1)=W_{k}(t)+\eta_{ckh}C_{kh}(t)-\frac{1}{\eta_{dkh}}D_{kh}(t)
\label{A3}
\end{equation}
\begin{equation}
B_{k,min}\leq B_{k}(t) \leq B_{k,max}, W_{k,min}\leq W_{k}(t) \leq W_{k,max}
\label{Wm}
\end{equation}
\begin{equation}
0\leq C_{ke}(t) \leq C_{ke,max}, 0\leq D_{ke}(t) \leq D_{ke,max}
\label{Cem}
\end{equation}
\begin{equation}
0\leq C_{kh}(t) \leq C_{kh,max}, 0\leq D_{kh}(t) \leq D_{kh,max}
\label{Chm}
\end{equation}
where $B_{k}(t)$, $\eta_{cke}$, $C_{ke}(t)$, $\eta_{dke}$ and $D_{ke}(t)$ denote the stored electricity, charging efficiency, charging rate, discharging efficiency and discharging rate of the battery. $W_{k}(t)$, $\eta_{ckh}$, $C_{kh}(t)$, $\eta_{dkh}$ and $D_{kh}(t)$ denote the stored thermal energy, charging efficiency, charging rate, discharging efficiency and discharging rate of the water tank.



\subsubsection{CHP}
A CHP unit generates heat and electricity simultaneously, which can be denoted by
\begin{equation}
E_{kCHP}(t)=\eta_{kpg}G_{kCHP}(t), H_{kCHP}(t)=\eta_{khg}G_{kCHP}(t)
\label{CHP1}
\end{equation}
\begin{equation}
0\leq E_{kCHP}(t) \leq E_{kCHP,max}, 0\leq H_{kCHP}(t) \leq H_{kCHP,max}
\label{CHP2}
\end{equation}
where $E_{kCHP}(t)$, $\eta_{kpg}$, $G_{kCHP}(t)$, $H_{kCHP}(t)$ and $\eta_{khg}$ are the electricity generation, electricity generation efficiency, natural gas consumption, heat generation and heat generation efficiency of the CHP unit at time slot $t$, respectively. 

\subsubsection{Boiler}
A boiler generates heat by consuming gas, which is 
\begin{equation}
H_{kb}(t)=\eta_{kbg}G_{kb}(t)
\label{bo1}
\end{equation}
\begin{equation}
0\leq H_{kb}(t) \leq H_{kb,max}
\label{bo2}
\end{equation}
where $H_{kb}(t)$, $\eta_{kbg}$ and $G_{kb}(t)$ are the heat generation, heat generation efficiency and natural gas consumption of the boiler, respectively. 

\subsection{Energy Trading with the Utility Companies}
For the industrial park, the electricity and natural gas purchased from the utility companies are denoted by $E(t)$ and $G(t)$, respectively. In addition, the park can sell redundant electricity $E_{o}(t)$ to the electricity utility company. There are maximum constraints of trading energy with the utility companies during a time slot.
\begin{equation}
0\leq E(t) \leq E_{max},  0\leq G(t) \leq G_{max} , 0\leq E_{o}(t) \leq E_{o,max}
\label{eo2}
\end{equation}

\subsection{Energy Balance}

The energy balance in the industrial park is 
\begin{subequations}
\begin{equation}
E_{tot}(t)=\sum_{k\in \boldsymbol{K}}E_{k}(t)+E(t)-E_{o}(t)
\label{11a}
\end{equation}
\begin{equation}
 G_{tot}(t)=G(t)-\sum_{k\in \boldsymbol{K}}[G_{kCHP}(t)+G_{kb}(t)]
 \label{11b}
 \end{equation}
\begin{equation}
H_{tot}(t)=\sum_{k\in \boldsymbol{K}}H_{k}(t)
\label{11c}
\end{equation}
\begin{equation}
E_{k}(t)=E_{kCHP}(t)+D_{ke}(t)-C_{ke}(t)+R_{k}(t)
\label{11d}
\end{equation}
\begin{equation}
H_{k}(t)=H_{kCHP}(t)+H_{kb}(t)+D_{kh}(t)-C_{kh}(t)
\label{11e}
\end{equation}
\end{subequations}
where (\ref{11a}), (\ref{11b}) and (\ref{11c}) denote the balance of electricity, gas and heat, and (\ref{11d}) and (\ref{11e}) denote the electricity and heat supplied by MEGP $k$. $E_{tot}(t)$, $G_{tot}(t)$ and $H_{tot}(t)$ are the total available electricity, gas and heat, respectively. $E_{k}(t)$ and $H_{k}(t)$ are the electricity and heat generation of MEGP $k$. $R_{k}(t)$ is the renewable energy generation of MEGP $k$. 

 For each industrial user, electricity loads are divided into IL and EL. For simplicity, heat and gas loads are considered as EL. Suppose there are $\boldsymbol{I}=\{1, 2, ..., I\}$ users for IL and $\boldsymbol{Q}=\{1, 2, ..., Q\}$ types of EL. 
The available energy domain of MEGP $k$ can be denoted as:
\begin{equation}
0\leq x_{k}(t)\leq x_{k,max}
\label{eqdt1}
\end{equation}
\begin{equation}
x_{k}(t)=\sum_{i \in \boldsymbol{I}}x_{ki}(t)+\sum_{q\in \boldsymbol{Q}}x_{kq}(t)
\label{eqdt}
\end{equation}
where $x_{k}(t)$ is the total energy generation of MEGP $k$ for energy $x \in \boldsymbol X$ at time slot t,  and $\boldsymbol X = \{E, H, G\}$, i.e., the set of electricity, heat and gas.  $x_{ki}(t)$ is the amount of IL for user $i$ satisfied by MEGP $k$. $x_{kq}(t)$ is the amount of EL $q$ for users satisfied by MEGP $k$. 

Supposing that some electricity IL can be cut down when necessary.  When there are high energy demands or power outages, the industrial park offers incentive price $p_{i}(t)$ for user $i$ to reduce their electricity IL by an amount ${X}_{ir}(t)$ without compromising the basic needs. User $i$ sets its reduced electricity IL by solving the following problem
\begin{equation}
\nonumber
\max_{0\leq {X}_{ir}(t) \leq \eta X_{i}(t)} p_{i}(t){X}_{ir}(t)-a_{i}{X}^{2}_{ir}(t)
\end{equation}
where $a_{i}{X}^{2}_{ir}(t)$ is the unsatisfactory cost.
$\eta$ is the ratio of maximum load reduction, and ${X}_{i}(t)$ is the original electricity IL of user $i$. 
When the incentive price satisfies
\begin{equation}
0\leq p_{i}(t)\leq 2a_{i}\eta {X}_{i}(t)
\label{pit}
\end{equation}
the optimal solution can be obtained by ${X}_{ir}(t) =p_{i}(t)/2a_{i}$, which means that $p_{i}(t)$ and ${X}_{ir}(t)$ are linearly dependent for user $i$, and the linear coefficient is $2a_{i}$. 

When incentive price $p_{i}(t)$ is given, the actual electricity IL of user $i$ is denoted as
\begin{equation}
\sum_{k\in \boldsymbol{K}_{i}}{x}_{ki}(t) =X_{i}(t)-p_{i}(t)/2a_{i}
\label{ail}
\end{equation}
where $\boldsymbol{K}_{i} \subseteq \boldsymbol{K}$ denotes the set of MEGPs supplied to user $i$. 

\section{Energy Management Scheme}
The objective of energy management is to minimize the time-averaged energy cost, subject to energy constraints. For simplicity, all variables of randomness are collected into $\boldsymbol{r}(t)=\{R(t), X(t)\}$, and all variables of optimization are collected into $\boldsymbol{M}(t)$=\{$x_{ki}(t)$, $x_{kq}(t)$, $x_{k}(t)$, $D_{ke}(t)$, $C_{ke}(t)$, $D_{kh}(t)$, $C_{kh}(t)$, $E_{o}(t)$, $E(t)$, $G(t)$, $X_{ir}(t)$\}. The cost of the industrial park at $t$ is denoted by 
\begin{equation}
\begin{aligned}
\phi(t) &= E(t)p_{e}(t)+G(t)p_{g}(t)-E_{o}(t)p_{o}(t)\\
&+\sum_{i\in \boldsymbol{I}} [p_{i}(t){X}_{ir}(t)-U_{i}(t)]-\sum_{k\in \boldsymbol{K}}\sum_{q\in \boldsymbol{Q}}U_{kq}(t)
\end{aligned}
\label{eq3011}
\end{equation}
$p_{e}(t)$ and $p_{g}(t)$ are the prices of the industrial park purchasing energy from the electricity and gas utility companies, respectively. $p_{o}(t)$ is the price of the industrial park selling energy to the electricity utility company. $U_{i}(t)$ is the satisfaction revenue of IL for user $i$, and $U_{kq}(t)$ is the satisfaction revenue from EL $q$ supplied by MEGP $k$.

The optimization problem of the industrial park is to find a scheduling policy to minimize the time-averaged cost, which can be expressed as a long-term optimization problem:
\begin{align}
\max_{\boldsymbol{M}(t)}\lim_{T\rightarrow\infty}\frac{1}{T}\sum_{t=0}^{T-1}\mathbb{E}\{\phi(t)\}
\label{eq301}
\end{align}
\begin{equation}
\nonumber
\text{s.t. } (\ref{A1}) - (\ref{ail})
\end{equation}
where the expectation considers all random variables.
However, the battery and water tank dynamics in (\ref{A1}) and (\ref{A3}) couple the optimization variables, which is not directly solved in most case. In addition, the knowledge of $\boldsymbol{r}(t)$ is causal, it is generally intractable to solve the optimization problem with the coupling across time. To solve the problem, it is essential to relax the time-coupling constraints (\ref{A1}) and (\ref{A3}), that is, replace them with average constraints. 

\subsection{Stochastic Energy Optimization}
Combining (\ref{A1})-(\ref{Wm}), the average energy charging and discharging in the long term should satisfy the following conditions
\begin{subequations}
\begin{equation}
\lim_{T\rightarrow\infty}\frac{1}{T}\sum_{t=0}^{T-1}\mathbb{E}\{C_{ke}(t)\}=\lim_{T\rightarrow\infty}\frac{1}{T}\sum_{t=0}^{T-1}\mathbb{E}\{D_{ke}(t)\}
\end{equation}
\begin{equation}
\lim_{T\rightarrow\infty}\frac{1}{T}\sum_{t=0}^{T-1}\mathbb{E}\{C_{kh}(t)\} =\lim_{T\rightarrow\infty}\frac{1}{T}\sum_{t=0}^{T-1}\mathbb{E}\{D_{kh}(t)\}
\end{equation}
\label{sto1}
\end{subequations}

In this way, (\ref{sto1}) ensures that the electricity charged equals the electricity discharged for a long period. The same situation is applied to the water tank. According to (\ref{sto1}), a relaxed version of (\ref{eq301}) is denoted by
\begin{equation}
\begin{aligned}
\max_{\boldsymbol{M}(t)}\lim_{T\rightarrow\infty}\frac{1}{T}\sum_{t=0}^{T-1}\mathbb{E}\{\phi(t)\}
\end{aligned}
\label{eq31}
\end{equation}
\begin{equation}
\nonumber
\text{s.t. } (\ref{Cem}) - (\ref{ail}), (\ref{sto1}) 
\end{equation}

To handle the coupling introduced by (\ref{sto1}), 
Lagrange multipliers $\lambda_{ke}$ and $\lambda_{kh}$ are introduced to associate with (\ref{sto1}). The Lagrangian function of (\ref{eq31}) is 
\begin{equation}
\begin{aligned}
L(\boldsymbol{M}(t), \boldsymbol{\lambda})&=\mathbb{E}\{\phi(t)\}+\sum_{k\in \boldsymbol{K}}\mathbb{E}[\lambda_{ke}(C_{ke}(t)-D_{ke}(t))]\\&+\sum_{k\in \boldsymbol{K}}\mathbb{E}[\lambda_{kh}(C_{kh}(t)-D_{kh}(t))]
\end{aligned}
\label{micro11}
\end{equation}

Its Lagrange dual function is 
\begin{equation}
\begin{aligned}
\Gamma(\boldsymbol{\lambda})&=\min_{\boldsymbol{M}(t)\in \widetilde {\boldsymbol{M}}(t)}L(\boldsymbol{M}(t), \boldsymbol{\lambda})
\end{aligned}
\label{micro12}
\end{equation}
where  $\widetilde {\boldsymbol{M}}(t)$ expresses the feasible set defined by the constraints (\ref{Cem}) - (\ref{ail}). The dual problem of (\ref{eq31}) can be denoted by 
\begin{equation}
\begin{aligned}
\max_{\boldsymbol{\lambda}}\Gamma(\boldsymbol{\lambda})
\end{aligned}
\label{micro13}
\end{equation}

For dual problem (\ref{micro13}), a gradient algorithm can be applied to get the optimal $\boldsymbol{\lambda}^{*}$. That is, the multipliers $\boldsymbol{\lambda}(t+1)$ at iteration $t+1$ are denoted as 
\begin{subequations}
\begin{equation}
\lambda_{ke}(t+1)=\lambda_{ke}(t)+\rho (C_{ke}(t)-D_{ke}(t))
\end{equation}
\begin{equation}
\lambda_{kh}(t+1)=\lambda_{kh}(t)+\rho (C_{kh}(t)-D_{kh}(t))
\end{equation}
\label{micro16}
\end{subequations}
\\where $C_{ke}(t)$, $D_{ke}(t)$, $C_{kh}(t)$ and $D_{kh}(t)$ are obtained by solving 
\begin{equation}
\begin{aligned}
 \Phi^*&=\min_{M(t)} \Phi(t)=\min_{M(t)} \phi(t)+\sum_{k\in \boldsymbol{K}}[\lambda_{ke}(t)(C_{ke}(t)\\&-D_{ke}(t))+\lambda_{kh}(t)(C_{kh}(t)-D_{kh}(t))]
\end{aligned}
\label{micro17}
\end{equation}
\begin{equation}
\nonumber
\text{s.t. } (\ref{Cem}) - (\ref{ail})
\end{equation}

Owing to the dual decomposition of optimization variables across time, the stochastic iterations are feasible. There are two additional benefits for the stochastic iterations in (\ref{micro16}) and (\ref{micro17}). Firstly, the solution for (\ref{micro17}) can be made close to the solution for (\ref{eq31}). Secondly, the solution for (\ref{micro17}) is feasible to the original problem (\ref{eq301}) when properly initialized. 

\subsection{Distributed Fast Implementation}

To obtain the optimal solution and lower computational complexity, dual decomposition methods are used again to dualize the constraint (\ref{eqdt}) in (\ref{micro17}), which couples EL and IL. The IL must be served per time instant $t$, which means the distributed algorithm must run some iterations at micro-slots per time instant. Thus, there are two timescales for the optimization algorithm.  The instantaneous Lagrange function can be denoted as 
\begin{equation}
\begin{aligned}
\overline{L}(\boldsymbol{M}, \boldsymbol{\tau}): =\Phi(t)+\sum_{k\in \boldsymbol{K}}\tau_{k}(\sum_{i\in \boldsymbol{I}}x_{ki}+\sum_{q\in \boldsymbol{Q}}x_{kq}-x_{k})
\end{aligned}
\label{micro18}
\end{equation}
where $\boldsymbol{\tau}=\{\tau_{1}, ..., \tau_{K}\}$ are the corresponding instantaneous Lagrange multipliers, and dual variable $\boldsymbol{\lambda}(t)$ in $\Phi(t)$ is updated in a slow time scale, which means that $\boldsymbol{\lambda}(t)$ can be treated as constants when solving the problem (\ref{micro17}). The dual function of (\ref{micro17}) is denoted by 
\begin{equation}
\overline{\Gamma}(\boldsymbol{\tau}) :=\min_{\boldsymbol{M}\in \overline{\boldsymbol{M}}}\overline{L}(\boldsymbol{M}, \boldsymbol{\tau})
\label{micro19}
\end{equation}
where $\overline{\boldsymbol{M}}$ is the feasible set subject to the constraints (\ref{Cem}) - (\ref{eqdt}). Then, the dual problem of (\ref{micro17}) is denoted as 
\begin{equation}
\max_{\boldsymbol{\tau}} \overline{\Gamma}(\boldsymbol{\tau})
\label{micro20}
\end{equation}

Different from the problem (\ref{micro13}),  which aims to achieve the stochastic estimate, 
the goal of (\ref{micro20}) is to make each user and each MEGP schedule its energy in a fully distributed way. In the following sections, two different gradient methods are proposed to solve the problem (\ref{micro20}).

First, the dual gradient algorithm is used to obtain Lagrange multipliers $\boldsymbol{\tau}$, and the iteration of $\boldsymbol{\tau}$ can be denoted as 
\begin{equation}
\boldsymbol{\tau}(n+1)=\boldsymbol{\tau}(n)+\sigma\nabla \overline{\Gamma}(\boldsymbol{\tau}(n))
\label{micro21}
\end{equation}
where $n$ is the iteration index of the micro-slot, and $\sigma$ is the stepsize. Then, the gradient for $\boldsymbol{\tau}(n)$ is denoted by
\begin{equation}
\nabla \overline{\Gamma}(\tau_{i}(n))=\sum_{i\in \boldsymbol{I}}x_{ki}(n)+\sum_{q\in \boldsymbol{Q}}x_{kq}(n)-x_{k}(n)
\label{micro22}
\end{equation}

Based on ($\ref{micro19}$), the industrial park can acquire an energy allocation $\boldsymbol{M}(n)$ by solving 
\begin{equation}
\begin{aligned}
\min_{\boldsymbol{M}(n)} & \, E(n)p_{e}(n)+G(n)p_{g}(n)-E_{o}(n)p_{o}(n)+\lambda_{ke}(n)(C_{ke}(n)\\&-D_{ke}(n))+\lambda_{kh}(n)(C_{kh}(n)-D_{kh}(n))-\tau_{k}(n)x_{k}(n)\\&+\sum_{k\in \boldsymbol{K}_{i}}\tau_{k}(n)x_{ki}(n)+p_{i}(n){X}_{ir}(n)-U_{i}(n)\\&+ \tau_{k}(n)x_{kq}(n)-U_{kq}(n)
\end{aligned}
\label{micro23}
\end{equation}
\begin{equation}
\nonumber
\text{s.t. } (\ref{Cem}) - (\ref{eqdt1}),  (\ref{pit}) - (\ref{ail})
\end{equation}
where ${X}_{ir}(n)=p_{i}(n)/2a_{i}$ according to Section II. Since (\ref{micro17}) is convex and satisfies the Slater condition, the solution of the dual problem (\ref{micro20}) yields the solution of the original problem (\ref{micro17}). When constant $\sigma$ is selected, gradient iterations (\ref{micro21}) will converge to a neighborhood of $\boldsymbol{\tau}^{*}$, and the value of the primal variables will be made tight approximation to the optimal solution \cite{Bertsekas2009Convex}. 

Although the dual gradient algorithm is used widely, its convergence rate is slow. To solve the issue in real time, a fast scheme based on the fast iterative shrinkage-thresholding algorithm (FISTA) \cite{Beck2009A}, which realizes fast convergence than that of the dual gradient algorithm, is proposed.

Different from the dual gradient algorithm in (\ref{micro21}), 
the fast scheme introduces an iteration $\overline{\boldsymbol{\tau}}(n)$ by combining the two recent iterations $\boldsymbol{\tau}(n)$ and $\boldsymbol{\tau}(n-1)$, which is denoted by
\begin{equation}
\overline{\boldsymbol{\tau}}(n)=(1-\epsilon)\boldsymbol{\tau}(n)+\epsilon \boldsymbol{\tau}(n-1)
\label{micro25}
\end{equation}
where $\epsilon=(1-\theta_{\tau}(n-1))/\theta_{\tau}(n)$, and the weight $\theta_{\tau}(n)$ is updated by
\begin{equation}
\theta_{\tau}(n)=(1+\sqrt{1+4\theta^{2}_{\tau}(n-1)})/2
\label{micro26}
\end{equation}

The Lagrange multiplier $\boldsymbol{\tau}(n)$ is obtained by a gradient ascent iteration based on $\overline{\boldsymbol{\tau}}(n)$
\begin{equation}
\tau_{k}(n+1)=\overline{\tau}_{k}(n)+\sigma(\sum_{i\in \boldsymbol{I}}x_{ki}(n)+\sum_{q\in \boldsymbol{Q}}x_{kq}(n)-x_{k}(n))
\label{micro27}
\end{equation}

The implementation process of fast scheme is on display in Algorithm 1. By using the two recent iterations, the ``combined'' iteration $\overline{\tau}(n)$ reduces the undesirable fluctuation of the gradient ascent iteration to achieve convergence. 


\begin{algorithm}[h]
\floatname{algorithm}{Algorithm}
\renewcommand{\algorithmicrequire}{\textbf{Communication}}
\renewcommand{\algorithmicensure}{\textbf{Contribution}}
\footnotesize
\caption{: Fast Scheme}
\label{alg1}
\begin{algorithmic}[1]
  \State \textbf{Initialize:} $\boldsymbol{\tau}(0)$, $\boldsymbol{\tau}(1)$, weight $\theta_{\tau}(0)$ and stepsize $\sigma$.   
    \For{$n = 1, 2, ...$}
\State Update $\theta_{\tau}(n)$ according to (\ref{micro26}).
\State Perform $\overline{\boldsymbol{\tau}}(n)$ based on (\ref{micro25}).
\State MEGPs send $\overline{\boldsymbol{\tau}}(n)$ to users.
\State Each MEGP and user calculate the solution $\boldsymbol{M}(n)$ using $\boldsymbol{\tau}=\overline{\boldsymbol{\tau}}$ based on (\ref{micro23}). 
\State Update $\boldsymbol{\tau}(n+1)$ according to (\ref{micro27}).
\EndFor 
\label{code:recentEnd}
\end{algorithmic}
\end{algorithm}

Then, the distributed stochastic gradient algorithm proposed in this paper integrates the dual gradient algorithm and the fast scheme, which is on display in Algorithm 2.
\begin{algorithm}[h]
\floatname{algorithm}{Algorithm}
\renewcommand{\algorithmicrequire}{\textbf{Communication}}
\renewcommand{\algorithmicensure}{\textbf{Contribution}}
\footnotesize
\caption{: Distributed Stochastic Gradient Algorithm}
\label{alg2}
\begin{algorithmic}[1]
  \State \textbf{Initialize:} $\boldsymbol{\lambda}(0)$, and stepsize $\rho$.   
      \For{$t= 1, 2, ...$}
\State Obtain $\boldsymbol{\tau}^{*}(t)$ according to (\ref{micro21}) or Algorithm 1.
\State Perform 
$\boldsymbol{M}(t)$ using $\boldsymbol{\tau}^{*}(t)$ by solving (\ref{micro23}).
\State Based on the solution $C_{ke}(t)$, $D_{ke}(t)$, $C_{kh}(t)$ and $D_{kh}(t)$, MEGP $k$ updates $\lambda_{ke}(t+1)$ and $\lambda_{kh}(t+1)$ according to (\ref{micro16}).      
\EndFor 
\label{code:recentEnd}
\end{algorithmic}
\end{algorithm}

Due to limited space, the performance analysis is not given here.

\section{Simulation Result}

In this section, the numerical results based on realistic data are presented to evaluate the proposed algorithm.
\subsection{Setup}
\begin{figure}
\centering
\begin{minipage}{1\linewidth}
  \centerline{\includegraphics[width=\hsize]{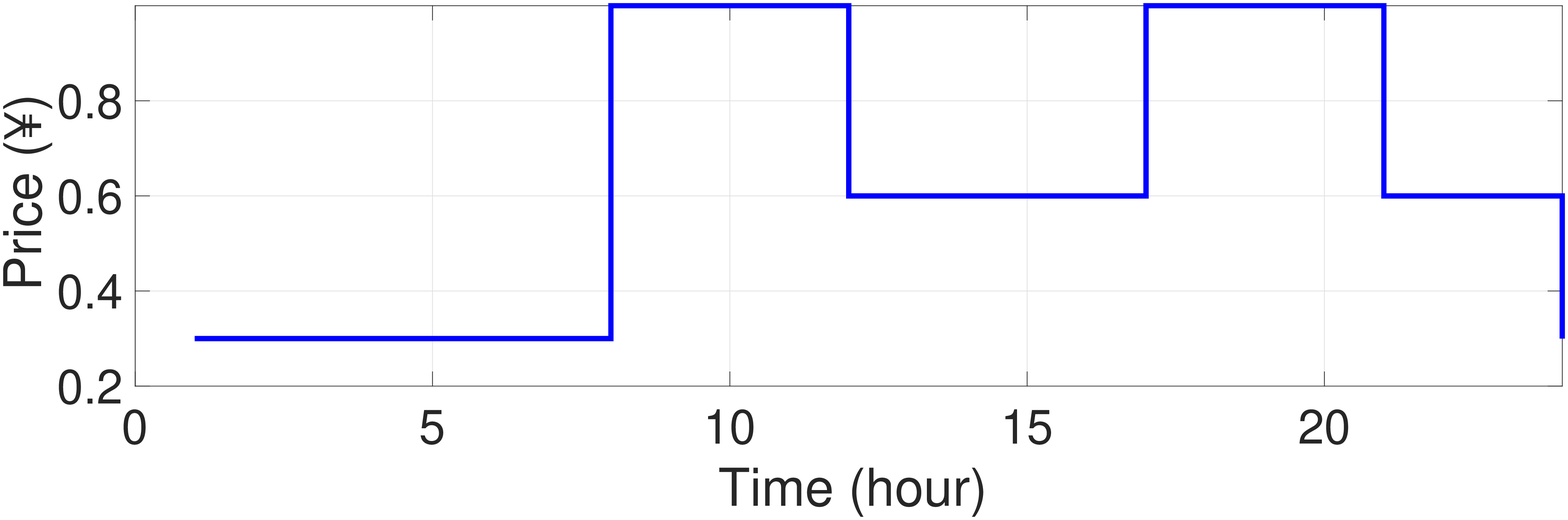}}
  \centerline{\scriptsize{(a) }}
\end{minipage}
\hspace{-5pt}
\begin{minipage}{1\linewidth}
  \centerline{\includegraphics[width=\hsize]{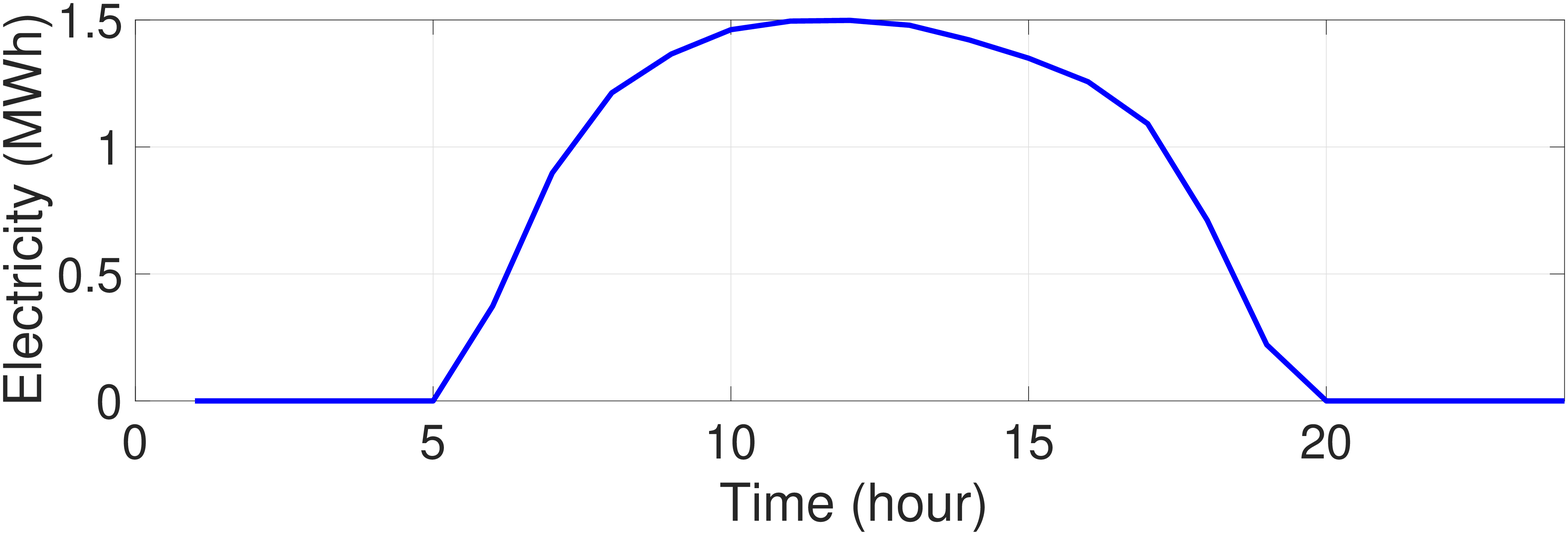}}
  \centerline{\scriptsize{(b) }}
\end{minipage}
\begin{minipage}{1\linewidth}
  \centerline{\includegraphics[width=\hsize]{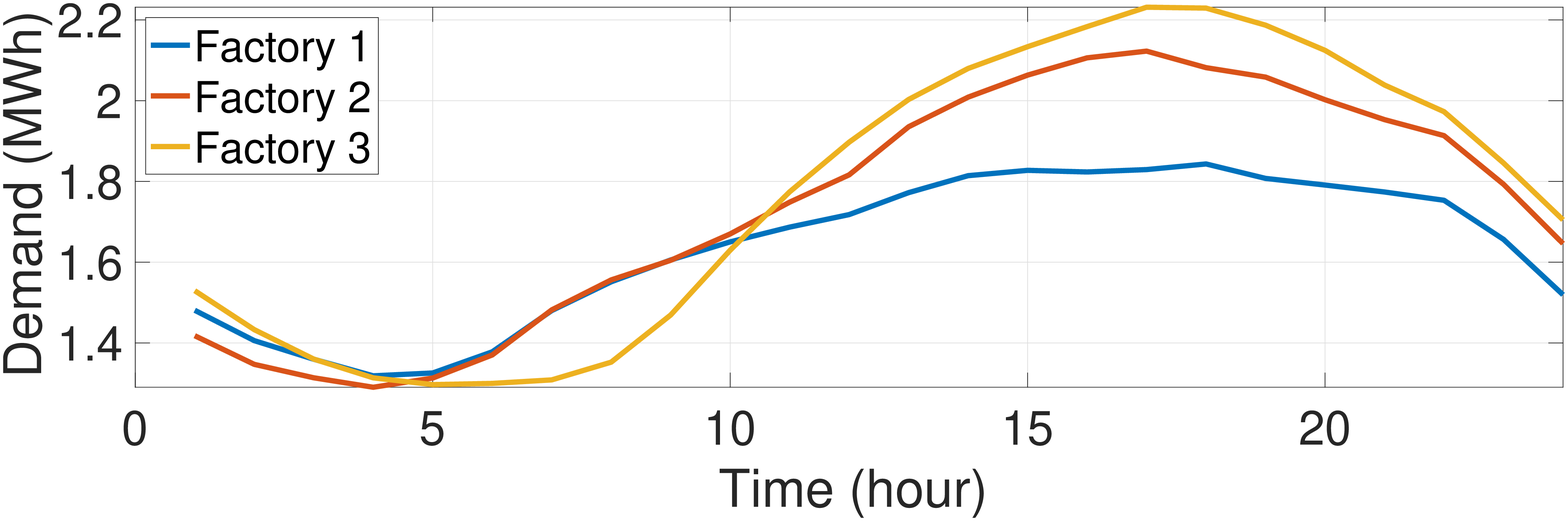}}
  \centerline{\scriptsize{(c) }}
\end{minipage}
\caption{Data of the industrial park. (a) Electricity Price. (b) Renewable Energy. (c) Electricity demand. }
\label{fig2}
\end{figure}

An industrial park consisting of two MEGPs and three factories is considered. 
The ratio of maximum electricity IL reduction is $\eta=0.15$. The unsatisfactory coefficient is $a_{i}=1$, and the price of gas is  $p_{g}(t)=0.4$ \textyen/kWh. 
Other related parameters are listed in Table I. 
 To verify the performance of the proposed management framework, two cases were adopted for comparison: Case 1 is based on a gradient method without incentive price, and all electricity loads are ILs. Case 2 is a scenario without renewable energy, based on dual gradient algorithm. To simulate real industrial scenes,
the price provided by the JiangSu Electric Power Company  \cite{Illinois} is shown in Fig.~\ref{fig2} (a). 
The data of photovoltaic systems provided by Renewables.ninja \cite{Institute} is shown in Fig.~\ref{fig2} (b). The hourly load provided by PJM hourly load \cite{PJM} is shown in Fig.~\ref{fig2} (c). 

\begin{table}
\centering
\caption{Related Parameters}
\begin{tabular}{|l |l |l |}
\hline
$\eta_{kpg}$, $\eta_{khg}$ & $\eta_{kbg}$ & $\eta_{cke}$, $\eta_{dke}$, $\eta_{ckh}$, $\eta_{dkh}$\\
\hline
35\% & 80\% & 98\%  \\
\hline
\end{tabular}
\\ [5pt]
\begin{tabular}{|l |l |l |}
\hline
$B_{k,max}, W_{k,max}$ & $C_{ke,max}$, $D_{ke,max}$ & $C_{kh,max}$, $D_{kh,max}$ \\
\hline
4MWh & 1MWh & 1MWh  \\
\hline
\end{tabular}
\label{TAB1}
\end{table}


\subsection{Performance Verification}

\begin{figure}
\centering
\begin{minipage}{1\linewidth}
  \centerline{\includegraphics[width=\hsize]{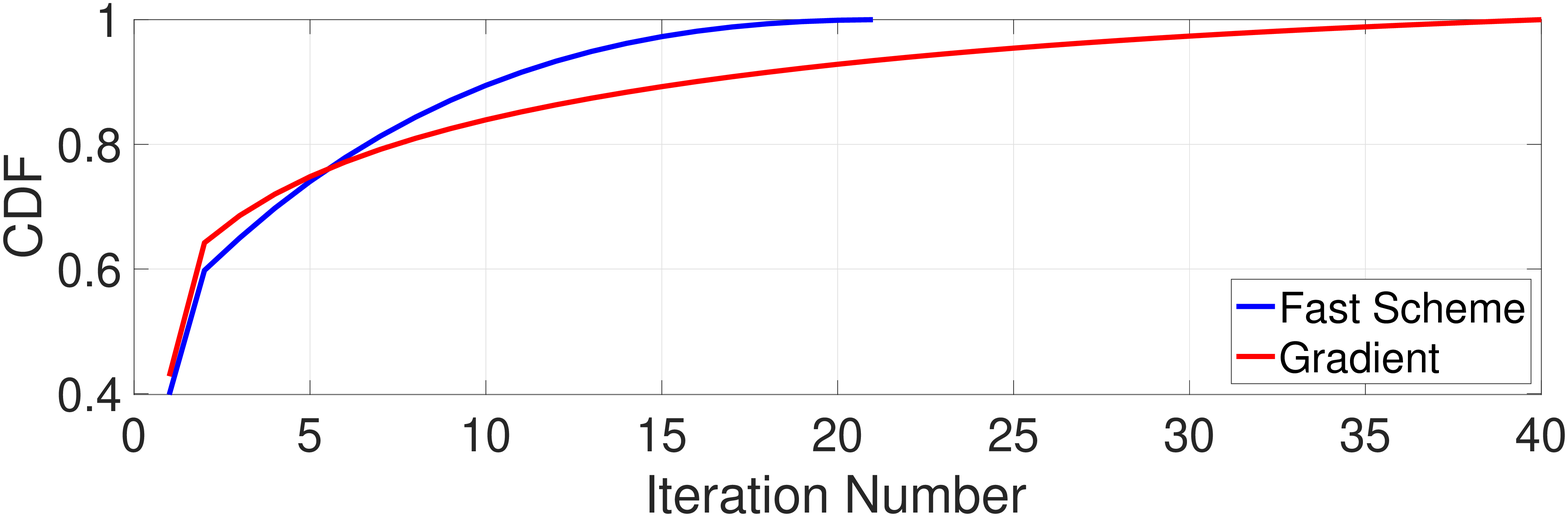}}
  \centerline{\scriptsize{(a) }}
\end{minipage}
\begin{minipage}{1\linewidth}
  \centerline{\includegraphics[width=\hsize]{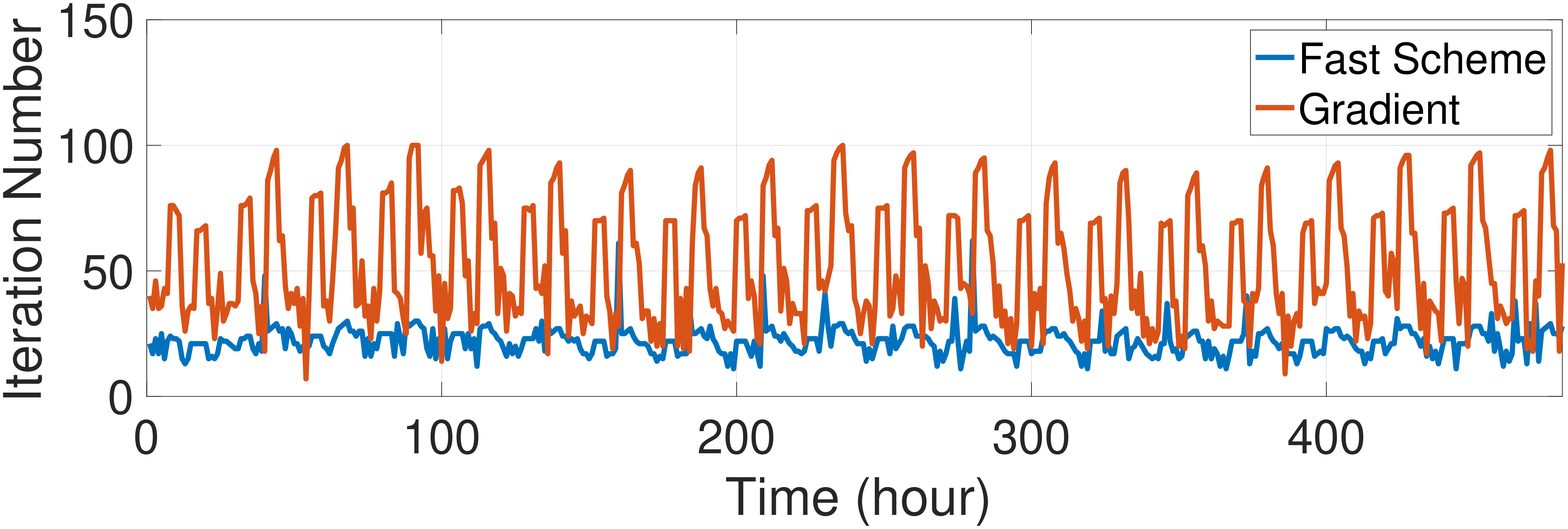}}
  \centerline{\scriptsize{(b) }}
\end{minipage}
\caption{Comparison for different methods. (a) CDF of iteration number. (b) Iteration number across 480 time slots.}
\label{fig3}
\end{figure}

Fig.~\ref{fig3} (a) and Fig.~\ref{fig3} (b) shows the convergence situation of the solution for problem (\ref{micro17}).  Fig.~\ref{fig3} (a) shows the cumulative distribution function (CDF) of iteration number for convergence at $t=1$. To show the comparison clearly, the simulation time is extended to $T=480$ slots, each slot is considered with 100 micro-slots at most. When the iteration number for (\ref{micro17}) is more than 100 or the difference of two iterations is less than 0.01, the iterations are stopped. The iteration stepsizes of the gradient algorithm (\ref{micro21}) and the fast scheme (\ref{micro27}) are both $\sigma=0.2$. Fig.~\ref{fig3} (b) shows that the fast scheme converges in about 20 iterations, and the dual gradient algorithm converges in more than 40 iterations mostly. 

\begin{figure}
\centering
\begin{minipage}{1\linewidth}
  \centerline{\includegraphics[width=\hsize]{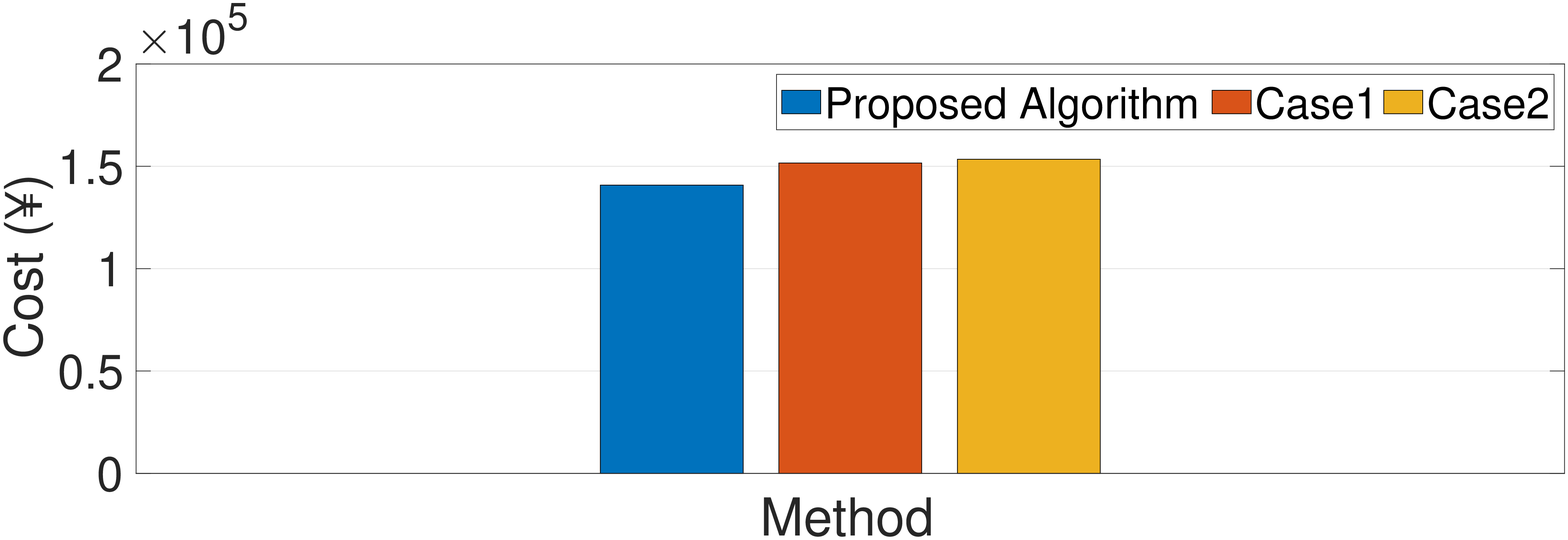}}
  \centerline{\scriptsize{(a) }}
\end{minipage}
\begin{minipage}{1\linewidth}
  \centerline{\includegraphics[width=\hsize]{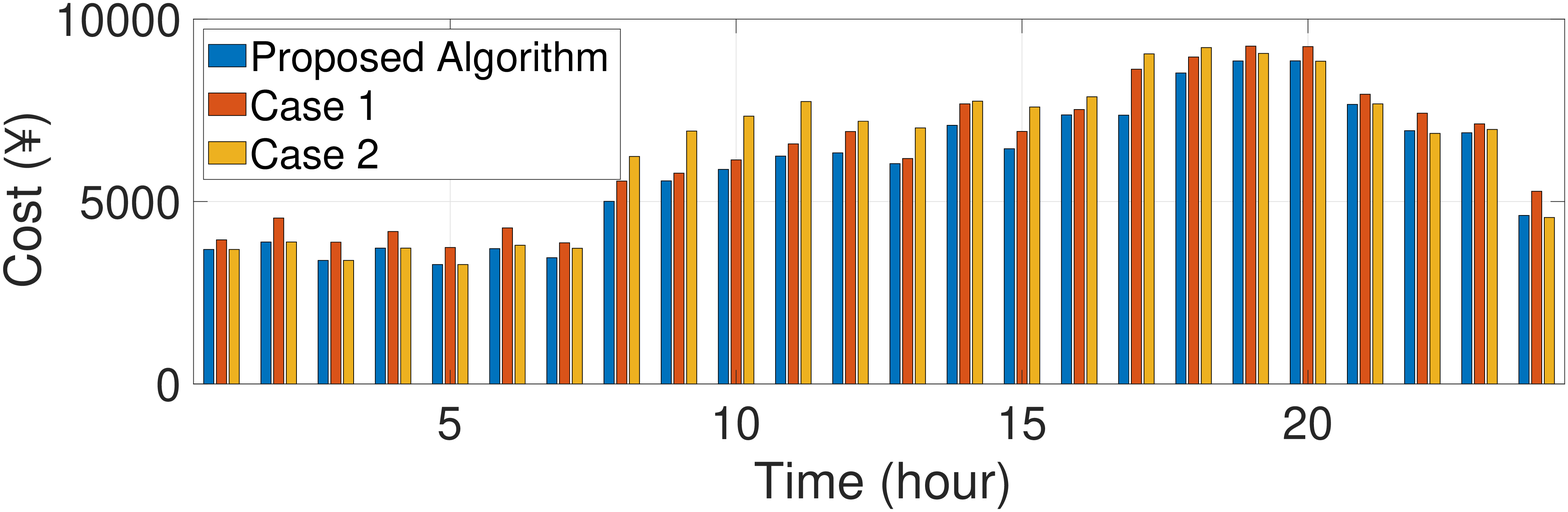}}
  \centerline{\scriptsize{(b) }}
\end{minipage}
\caption{Comparison of costs for different cases. (a) Total costs. (b) Costs across 24 time slots.}
\label{fig4}
\end{figure}

Fig.~\ref{fig4} (a) and Fig.~\ref{fig4} (b) show the total costs and the costs across 24 time slots of the proposed algorithm, Case 1 and Case 2. Case 1, which has no ELs and incentive revenue, is sensitive to the variations of renewable energy, prices and loads. Thus, the cost of Case 1 is higher than that of proposed algorithm all day. Case 2, which has no renewable energy, is sensitive to the energy price and needs to buy more energy in daytime, and is close to the proposed algorithm at night. Both of Case 1 and Case 2 incur high costs. Thus, the proposed algorithm makes full use of IL, EL and incentive revenue to schedule energy flexibly, and leverages renewable energy and energy storage to reduce the cost in real time. 


\begin{figure}
\centering
\begin{minipage}{1\linewidth}
  \centerline{\includegraphics[width=\hsize]{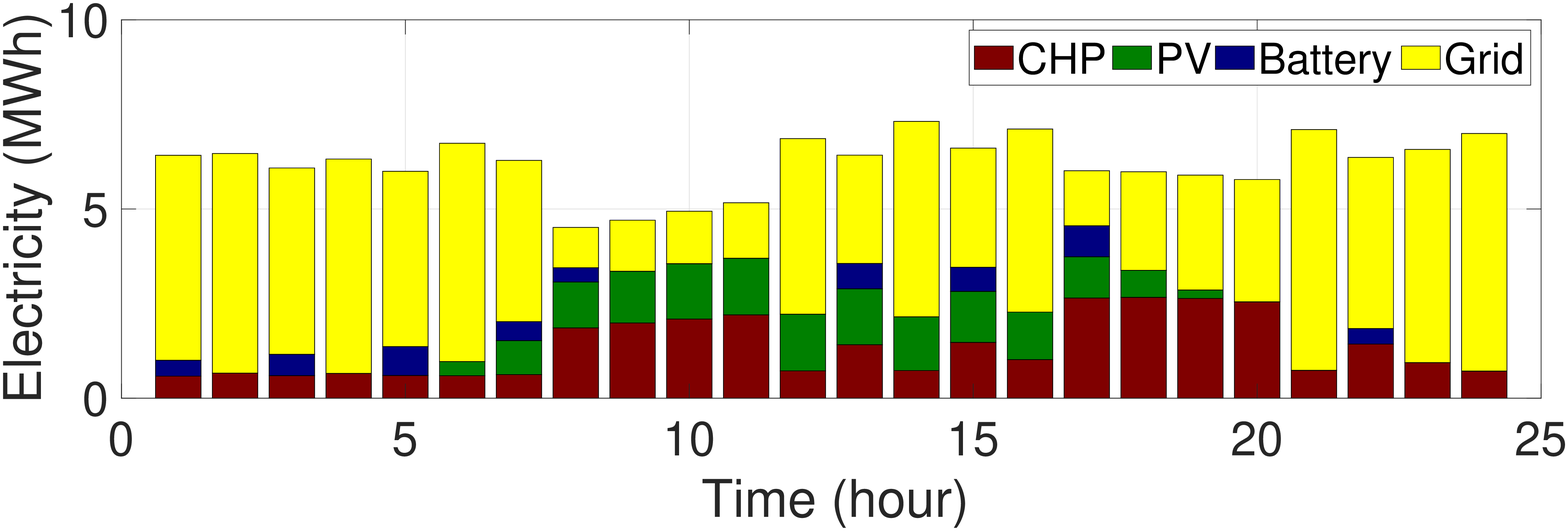}}
  \centerline{\scriptsize{(a) }}
\end{minipage}
\begin{minipage}{1\linewidth}
  \centerline{\includegraphics[width=\hsize]{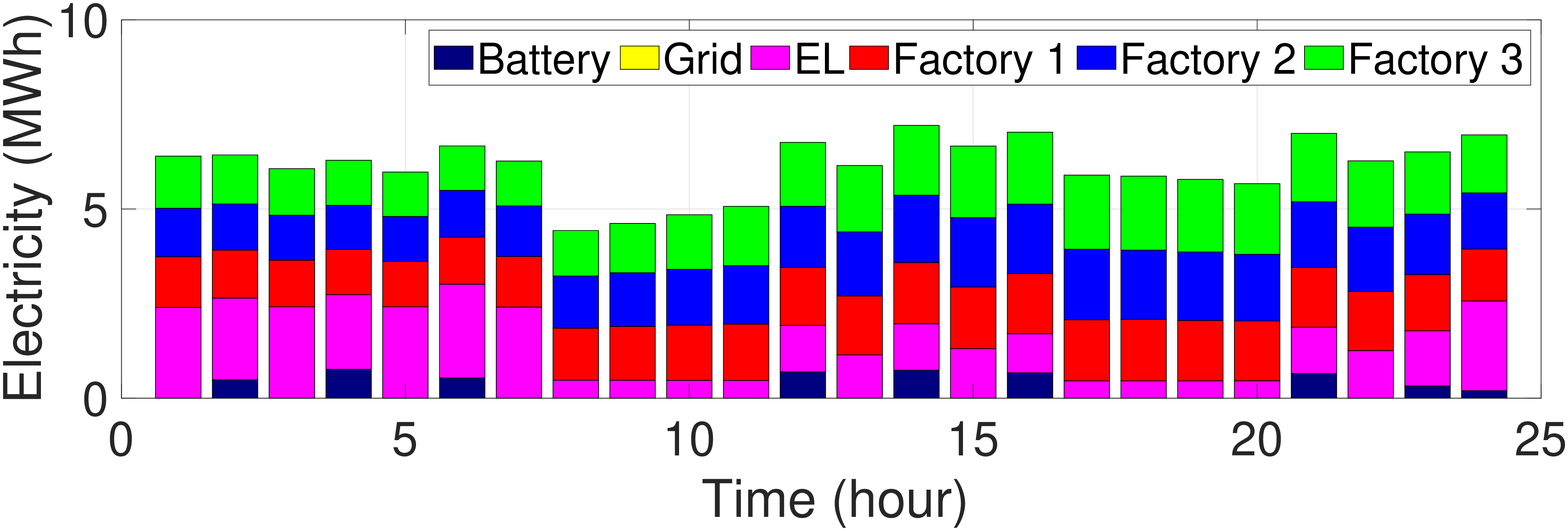}}
  \centerline{\scriptsize{(b) }}
\end{minipage}
\caption{Electricity profiles of the industrial park using the proposed algorithm. (a) Electricity generation. (b) Electricity consumption.}
\label{fig5}
\end{figure}

The energy generation and consumption of the park are shown in Fig.~\ref{fig5}. Since the electricity price during 8:00-23:00 is higher than other time, the park uses the CHP units to generate more electricity instead of purchasing electricity at a high price and simultaneously generate thermal energy for the heat demand, which is shown in Fig.~\ref{fig5}. The batteries are charged at low-price time 12:00, 21:00 and 24:00, and discharged at high-price time 8:00 and 17:00, which is on display in Fig.~\ref{fig5}. Fig.~\ref{fig5} shows that the EL is smaller when the electricity price is high. When the electricity price is not fixed, a similar effect can be obtained. Therefore, the proposed algorithm realize the energy demand shift, energy trading with the electricity, and energy supply by energy storage flexibly. 

\section{Conclusion}
In this paper, a multi-energy management framework in an industrial park including multi-energy generation plants and users was presented, which fully mobilizes the coordination mechanism of energy supply, demand and storage. The energy demand management problem was constructed as a long-term optimization problem to further consider the spatiotemporal variation of loads, renewables and energy prices. A distributed stochastic gradient algorithm based on dual decomposition and a fast scheme were proposed to reduce the influence of multi-energy coupling, time-varying renewable energy generation and demand. 
At last, the simulation results based on realistic data verified the performance of the proposed mechanisms. 

%
%
%

\end{document}